\documentclass[aps,pre,twocolumn,amsfonts,longbibliography,citeautoscript,superscriptaddress,10pt]{revtex4-2}
\usepackage{xcolor}
\usepackage{bm}
\usepackage{bbold}
\usepackage{amsmath}
\usepackage{amssymb}
\usepackage{epsfig}
\usepackage{verbatim}
\usepackage[T1]{fontenc}
\usepackage{array}

\newcommand{\im}{\mathrm{Im}}
\newcommand{\avC}[1]{\left \langle #1 \right\rangle}
\newcommand{\exValue}[1]{\langle #1 \rangle}

\newcommand{\citetemp}[1]{[\textbf{pending}]}

\DeclareMathOperator*{\extr}{extr}

\usepackage[colorlinks=true,bookmarks=false,citecolor=blue,linkcolor=blue,hyperfootnotes=true,urlcolor=blue]{hyperref}

\begin{document}

\title{Replica theory for the rate functional of the empirical spectral distribution function of diluted Hermitian matrices}
\author{Edgar Guzm\'an-Gonz\'alez}
\email{edgar.guzman@hainanu.edu.cn}
\affiliation{School of Physics and Optoelectronic
Engineering, Hainan University, 570228 Haikou, P. R. China}

\author{Isaac P\'erez Castillo}
\email{iperez@izt.uam.mx}
\affiliation{Departamento de F\'isica, Universidad Aut\'onoma Metropolitana-Iztapalapa, San Rafael Atlixco 186, Ciudad de M\'exico 09340, M\'exico}
\date{\today}

\begin{abstract}
We develop a replica-based framework for the scaled cumulant-generating functional of the empirical spectral distribution function $i_C$ of diluted Hermitian random matrices. Within a replica-symmetric saddle-point assumption, this construction yields a candidate rate functional for fluctuations of $i_C$. As an illustrative application, we consider adjacency matrices of unweighted Erd{\H{o}}s--R\'enyi random graphs with mean degree $c$. We derive explicit expressions for the first two cumulants of $i_C$, indicate how higher cumulants can be obtained from further functional derivatives, and compute the rate function of Fourier coefficients, equivalently of selected linear spectral statistics. The replica-symmetric predictions are tested against exact numerical diagonalization and show good agreement in the accessible fluctuation regime. The approach provides a basis for studying rate functionals of spectral observables in sparse random matrix ensembles.
\end{abstract}

\maketitle

{\it Introduction.---} Understanding the spectral properties of large complex systems is a central problem in many areas of physics and applied mathematics. Random matrix theory (RMT) provides a quantitative framework for describing universal features of such spectra, with origins in multivariate statistics \cite{wishart1928generalised} and nuclear physics \cite{wigner1955characteristic, dyson1962statistical, Oxford2015}. For the Wigner and the classical invariant Hermitian ensembles, global spectral statistics are well understood: the semicircle law and its extensions can be derived using moment methods, orthogonal-polynomial techniques, and Coulomb-gas methods \cite{pastur1972spectrum, mehta2004random, forrester2010loggases}. More recent advances in local universality and eigenvalue rigidity show that many of these spectral features extend far beyond the classical ensembles \cite{erdos2011universality, tao2012topics}.

Sparse Hermitian random matrices, which naturally arise as adjacency or Laplacian operators of random graphs, display a richer phenomenology than dense mean-field ensembles \cite{vanmieghem2010graph, biroli1999random}. Finite connectivity produces effects that are absent, or strongly suppressed, in dense ensembles: deviations from the semicircle law \cite{rodgers1988density,bauer2001random,khorunzhy2004spectrum,semerjian2002sparse}, localized eigenstates and mobility edges \cite{abou1973selfconsistent, fyodorov1991localization, mirlin1996transition}, delta peaks and localized states associated with finite graph components or rare local structures \cite{biroli1999random,Golinelli2003}, and nontrivial behavior of spectral edges and local laws in sparse graphs \cite{benaychGeorges2020spectralRadii,lee2018local}. These properties have direct implications for modeling real-world systems, where sparse random matrices appear prominently in ecosystem dynamics \cite{allesina2015predicting, grilli2016, gibbs2018} and neural networks \cite{rajan2006eigenvalue, ahmadian2015random}.

To characterize these spectral properties quantitatively, it is convenient to introduce observables that capture the global distribution of eigenvalues. A central example is the empirical spectral distribution function $i_C$, where $i_C(x)$ denotes the fraction of eigenvalues smaller than $x \in \mathbb{R}$. Many other spectral quantities, such as the spectral density in the sense of distributions, the number of eigenvalues inside an interval, and linear spectral statistics, can be obtained from $i_C$ or from suitable projections of it \cite{diaconis2001linear}. If $C$ is random, $i_C$ itself becomes a random function, making it natural to study its statistical properties, including typical fluctuations and large fluctuations of spectral observables. In ensembles where such probabilities decay exponentially with the matrix size, these fluctuations are naturally described by a large-deviation rate functional for $i_C$.

Rate functions have been used extensively to analyze rare events for finite-dimensional spectral observables, such as extreme eigenvalues \cite{dean2006large,majumdar2014top}. In dense Hermitian ensembles, rigorous large-deviation principles are available for empirical spectral measures or related spectral-density functionals \cite{benarous1997,hiai2000}. For sparse ensembles, the picture is more fragmented. Recent rigorous work has treated large deviations of edge or largest-eigenvalue observables in sparse random graphs and sparse Wigner-type matrices \cite{Bhattacharya2021spectralEdge,GangulyHiesmayrNam2024}, and a large-deviation principle for the empirical spectral measure has been established in a supercritical sparse Wigner regime with diverging mean degree \cite{Augeri2025esd}. Fluctuation theory for linear spectral statistics of random graph adjacency matrices has also advanced, including recent central-limit theorems for inhomogeneous random graphs across sparsity regimes \cite{ZhuZhu2025linearStats}. These results leave open the complementary constant-mean-degree setting addressed by cavity and replica methods, especially for functional fluctuations of the empirical spectral distribution function itself.

Analytical progress in the constant-connectivity regime requires methods capable of retaining the locally tree-like structure of the underlying graph in the thermodynamic limit. The replica method provides one such route, allowing spectral densities of sparse ensembles to be computed using tools from statistical mechanics \cite{edwards1976, mezard1988spin, mezardBook, mezard2001bethe, kuhn2008spectra, rogers2008cavity, Susca_2021}. On locally tree-like graphs, the same structure leads to cavity equations that relate spectral properties to recursive updates on the underlying graph and, in appropriate settings, to rigorous infinite-size limits \cite{bordenave2010resolvent}. Supersymmetric methods provide a complementary analytical framework for localization and related spectral phenomena \cite{efetov1997supersymmetry}. Previous replica calculations have also produced large-deviation rate functions for scalar spectral observables, such as the number of eigenvalues in an interval, and for related diluted covariance ensembles \cite{Metz2016, castillo2018large}.

In this work, we develop a replica-based framework for the scaled cumulant-generating functional of the empirical spectral distribution function $i_C$ in diluted Hermitian random matrices. For adjacency matrices of Erd{\H{o}}s--R\'enyi random graphs with fixed mean degree, the replica-symmetric saddle yields a candidate rate functional for fluctuations of $i_C$. We derive explicit expressions for the first two cumulants of $i_C$, indicate how higher cumulants can be obtained from further functional derivatives, and show how the same functional formalism gives rate functions for Fourier coefficients, equivalently for selected linear spectral statistics. The analytical predictions are compared with exact numerical diagonalization. The rest of the paper is organized as follows: we first define the model and the cumulant-generating functional, then derive the replica-symmetric saddle equations, extract cumulants and Fourier-mode rate functions, and finally test the predictions numerically.

\emph{Model definitions.---} Consider an ensemble of $N \times N$ real and symmetric matrices. For a given matrix $C$ in the ensemble, let $\lambda_1(C),\ldots,\lambda_N(C)$ denote its eigenvalues. We define the empirical spectral distribution function of $C$ as
\begin{equation}
i_C(x)=\frac{1}{N}\sum_{\alpha=1}^{N}\Theta\left(x-\lambda_\alpha(C)\right)\,,
  \label{eq:empiricalDistribution}
\end{equation}
so that, away from eigenvalues, $i_C(x)$ is the fraction of eigenvalues of $C$ smaller than $x$. At points where $x$ coincides with an eigenvalue, the value of $i_C$ is fixed by the Heaviside convention used in the logarithmic regularization below, corresponding to $\Theta(0)=1/2$. Thus an eigenvalue exactly equal to $x$ contributes one half to the sum in Eq.~\eqref{eq:empiricalDistribution}; this convention only affects the value of $i_C$ at jump discontinuities and not observables obtained by integration against integrable test functions.

Let $\mu$ be a real test function with compact support contained in a fixed bounded interval before the thermodynamic limit is taken. To study the statistics of the random function $i_C$, we introduce its cumulant-generating \emph{functional},
\begin{equation}
\begin{split}
F[\mu]= -\frac{1}{N}\ln\avC {e^{-N\int dx\, \mu(x) i_C (x)}}\,.
\end{split}
\label{eq:cgf}
\end{equation}
Here $\avC{(\cdots)}$ denotes the average over the ensemble of matrices $C$. The minus sign in Eq.~\eqref{eq:cgf} fixes the convention used throughout: $F$ is the negative scaled logarithm of the Laplace transform of $i_C$ with source $\mu$.

In the thermodynamic limit $N \to \infty$, we assume, consistently with the replica calculation developed below, that a large-deviation principle holds for the relevant values of the empirical spectral distribution function. Let $I \subset \mathbb R$ be a fixed bounded interval containing the support of $\mu$. We denote by $\zeta$ a candidate value of the random function $i_C$ restricted to $I$; equivalently, $\zeta(x)$ represents a possible empirical fraction of eigenvalues below $x$ for each $x\in I$. Denoting by $\mathbb{P}_N$ the probability law induced by the $N$-dimensional matrix ensemble, we write formally, for a sufficiently small neighborhood $\mathcal U_\zeta$ of $\zeta$,
\begin{equation}
\begin{split}
\mathbb{P}_N\left(i_C|_I\in \mathcal U_\zeta\right) \asymp e^{-N\Psi[\zeta]}\,,
  \end{split}
\end{equation}
up to subexponential factors at scale $N$. Here $\Psi[\zeta]$ denotes the candidate rate functional, and $\asymp$ denotes logarithmic equivalence. The neighborhood $\mathcal U_\zeta$ may be understood, at this formal level, with respect to a natural function-space topology on distribution functions, such as uniform or $L^1$ distance on $I$.

If $\Psi$ is a convex functional, the sign convention in Eq.~\eqref{eq:cgf} implies the functional Legendre transform \cite{Dembo2009,Touchette2009}
\begin{equation}
\begin{split}
\Psi[\zeta]&= -\extr\limits_{\mu} \left(\int dx\, \mu(x) \zeta(x)-F[\mu]\right)\\
&= \extr\limits_{\mu} \left(F[\mu]-\int dx\, \mu(x) \zeta(x)\right)\,,
\end{split}
\label{eq:legendreRate}
\end{equation}
where $\extr$ denotes the stationary (extremal) value with respect to $\mu$. In what follows, we compute the cumulant-generating functional for arbitrary compactly supported real $\mu$ using the replica method from spin-glass theory. The resulting replica-symmetric expression for $F[\mu]$ gives, through Eq.~\eqref{eq:legendreRate}, a candidate rate functional for fluctuations of $i_C$. Full details of the calculation are provided in the Supplemental Material, Sec.~\ref{app:asymmetric}.

\emph{Computing the cumulant-generating functional.---} First, we write $i_C$ in determinant form. Starting from the definition above and using $2\pi i\Theta(y)=\lim_{\eta\rightarrow0^+}[\ln(i\eta-y)-\ln(-i\eta-y)]$, we obtain
\begin{equation}
\begin{split}
e^{N i_C(x)} = \lim_{\eta \rightarrow 0^+}\left(\frac{\det[C-(x^\eta)^*\bm{1}]}{\det[C-x^\eta\bm{1}]}\right)^{\frac{1}{2\pi i}}\,,
\end{split}
\label{eq:enicx}
\end{equation}
where an asterisk denotes complex conjugation, $\bm{1}$ is the $N\times N$ identity matrix, and $x^\eta\equiv x+i\eta$. This sign convention is fixed by applying the identity above to $y=x-\lambda_i$.

To compute $F[\mu]$, we first partition an interval containing the support of $\mu$ and approximate the integral as $\int dx\,\mu(x)i_C(x)\approx \Delta x\sum_{j=1}^L\mu(x_j)i_C(x_j)$, where $x_0,\dots,x_L$ are the points in the partition and $\Delta x$ is the spacing. In the limit $\Delta x\rightarrow0$ this approximation becomes exact. After rewriting the determinants in Eq.~\eqref{eq:enicx} using Gaussian integrals, we obtain
\begin{equation}
\begin{split}
F[\mu]=& -\lim_{\Delta x\rightarrow0}\lim_{\eta\rightarrow0^+}\frac{1}{N} \ln \Bigg\langle \prod_{j=1}^L \left[\mathcal Z(x_j^\eta)\right]^{n_j} \left[\mathcal Z^*(x_j^\eta)\right]^{-n_j} \Bigg\rangle\,,
\end{split}
\label{eq:fmulimits}
\end{equation}
where
$n_j=\Delta x\,\mu(x_j)/(\pi i)$ and
\begin{equation}
\begin{split}
\mathcal Z(x^\eta)&=\frac{1}{(2\pi i)^{N/2}}\int dy^N\\
&\quad\times\exp\left\{\frac{i}{2}\sum_{m,n=1}^N y_m\left[C-(x^\eta)^*\bm{1}\right]_{mn}y_n\right\}.
\end{split}
\label{eq:Zxj}
\end{equation}
Equation~\eqref{eq:Zxj} is analogous to the partition function of a system with $N$ particles coupled through the matrix $C-(x^\eta)^*\bm{1}$. This mathematical analogy allows the use of tools from statistical physics to compute $F[\mu]$.

Using conventional methods, computing the average in Eq.~\eqref{eq:fmulimits} is generally nontrivial. We therefore employ the replica method from spin-glass theory to obtain an analytical expression for $F[\mu]$ in the thermodynamic limit. At the replica stage, the powers of $\mathcal Z$ and $\mathcal Z^*$ are introduced as independent integer replica numbers, say $n_j^+$ and $n_j^-$, and the averaged expression is analytically continued at the end to
$n_j^+=n_j$ and $n_j^-=-n_j$, with $n_j=\Delta x\,\mu(x_j)/(\pi i)$. This continuation is performed using a replica-symmetric ansatz \cite{Edwards, mezard1988spin}. The resulting expression should therefore be read as a replica-symmetric saddle-point prediction for the scaled cumulant-generating functional and, through Eq.~\eqref{eq:legendreRate}, for the associated candidate rate functional; its consequences are tested below by comparison with exact numerical diagonalization.

Up to this point, the approach is fully general. The success in applying it to compute $F[\mu]$ depends on the ensemble of matrices under consideration. As a nontrivial illustrative example, we consider adjacency matrices of Erd{\H{o}}s--R\'enyi random graphs. These are graphs in which two different nodes are connected with probability $c/N$, where $c>0$ is independent of $N$ and denotes the average connectivity of the graph. Specifically, the probability of drawing a matrix $C$ with components $c_{ij}$ in its upper triangular part is
\begin{equation}
P(C) = \prod_{i<j} \left[\left(1-\frac{c}{N}\right) \delta_{c_{ij},0} + \frac{c}{N} \delta_{c_{ij},1}\right]\,.
\label{eq:pPoisson}
\end{equation}
Beyond their analytical tractability, the Erd{\H{o}}s--R\'enyi random graphs \cite{Erdos1959} represent the simplest baseline model of random connectivity, against which the structure of more complex real-world systems---from ecosystems to neural architectures---is often compared \cite{newman2018networks, allesina2015predicting, rajan2006eigenvalue}.

As detailed in Supplemental Material \cite{SM}, our approach allows one to express the cumulant-generating functional $F[\mu]$ in terms of an effective theory defined over the space of complex-valued functions $\Delta$ with the same support as $\mu$. In this framework, $F[\mu]$ takes the following form,
\begin{equation}
F[\mu] = -\frac{c}{2} + \frac{1}{2}\int dx\, \mu(x) + S[\mu]\,,
\label{eq:FmuC}
\end{equation}
where $S[\mu]$ is a functional defined through path integrals over such functions $\Delta$ with normalized weight $w[\Delta]$. Specifically,
\begin{equation}
\begin{split}
S[\mu] &= \frac{c}{2}\int \mathcal D\Delta\mathcal D\Delta' w[\Delta]w[\Delta']\\
&\quad\times\exp\left\{\frac{1}{\pi i}\int dx\,\mu(x)\Lambda(\Delta'(x),\Delta(x))\right\}\\
&-\ln\Bigg\{\sum_{k=0}^\infty p_c(k)\int \mathcal D\Delta\prod_{r=1}^k\left[\mathcal D\Delta_r\,w[\Delta_r]\right]\\
&\hspace{1.0cm}\times\delta[\Delta-s_{\Delta_1,\dots,\Delta_k}]\\
&\hspace{1.0cm}\times\exp\left\{\frac{1}{2\pi i}\int dx\,\mu(x)\Xi(\Delta(x))\right\}\Bigg\}\,,
\end{split}
\label{eq:SmuC}
\end{equation}
where $p_c(k)=e^{-c}c^k/k!$ is the Poisson distribution with mean $c$, $\delta[\,\cdots\,]$ is a functional Dirac delta, and
\begin{equation}
s_{\Delta_1,\dots,\Delta_k}(x)=-\left(x-i\eta+\sum_{r=1}^k\Delta_r(x)\right)^{-1}\,.
\label{eq:sDeltaDef}
\end{equation}
We also introduced
\begin{equation}
\begin{split}
\Lambda(\Delta'(x),\Delta(x))&=-i\,\im\ln\left(1-\Delta(x)\Delta'(x)\right),\\
\Xi(\Delta(x))&=2i\,\im\ln\left(i\Delta(x)\right)\,.
\end{split}
\label{eq:functionDefinitions}
\end{equation}
The weight $w[\Delta]$ is determined self-consistently by
\begin{equation}
\begin{split}
w[\Delta]&=A\sum_{k=0}^\infty p_c(k)\int\prod_{r=1}^k\left[\mathcal D\Delta_r\,w[\Delta_r]\right]\\
&\quad\times\delta[\Delta-s_{\Delta_1,\dots,\Delta_k}]\\
&\quad\times\exp\left\{\frac{1}{2\pi i}\int dx\,\mu(x)\Xi(\Delta(x))\right\}\,,
\end{split}
\label{eq:wIf}
\end{equation}
where $A$ is fixed by the normalization $\int \mathcal D\Delta\,w[\Delta]=1$. With this normalization, Eqs.~\eqref{eq:FmuC}--\eqref{eq:wIf} give $S[0]=c/2$ and hence $F[0]=0$. Further derivations and intermediate steps are presented in \cite{SM}.

For observables depending on $i_C$ at a single spectral parameter $x$, the $\mu\equiv0$ recursion for $w$ closes at the level of the one-point marginal. Let $\pi_x(\omega)$ denote the marginal distribution of $\omega=\Delta(x)$ induced by $w[\Delta]$. Marginalizing Eq.~\eqref{eq:wIf} over all values of the function $\Delta$ away from $x$ gives
\begin{equation}
\begin{split}
\pi_x(\omega)&=\sum_{k=0}^\infty p_c(k) \int\prod_{r=1}^k\left[d\omega_r\,\pi_x(\omega_r)\right]\\
&\quad\times\delta\left(\omega+\left(x-i\eta+\sum_{r=1}^k\omega_r\right)^{-1}\right)\,.
\end{split}
\label{eq:fixedXMarginal}
\end{equation}
This is the standard replica saddle-point equation for diluted systems, written in the sign convention of Eq.~\eqref{eq:sDeltaDef}~\cite{kuhn2008spectra}.

From the previous expressions, several observables of physical interest can be obtained in a direct way. We focus on two representative consequences of the formalism: the first two cumulants of the empirical spectral distribution function $i_C$, and rate functions for Fourier projections of $i_C$, which are examples of linear spectral statistics.

By taking functional derivatives of $F[\mu]$ with respect to the source and then setting $\mu\equiv0$, we obtain the connected cumulants of $i_C$. Defining
\begin{equation}
\kappa_n(x_1,\ldots,x_n)=\avC{i_C(x_1)\cdots i_C(x_n)}_{\mathrm{c}}\,,
\end{equation}
the sign convention in Eq.~\eqref{eq:cgf} gives
\begin{equation}
\left.\frac{\delta^n F[\mu]}{\delta\mu(x_1)\cdots\delta\mu(x_n)}\right|_{\mu=0}=(-N)^{n-1}\kappa_n(x_1,\ldots,x_n)\,.
\label{eq:cumulantDerivative}
\end{equation}
In the explicit formulas below, $w[\Delta]$ denotes the solution of Eq.~\eqref{eq:wIf} at $\mu\equiv0$. For compactness, we write
\begin{equation}
\begin{split}
\left\langle A[\Delta]\right\rangle_w&=\int \mathcal D\Delta\,w[\Delta]A[\Delta]\,,\\
\left\langle B[\Delta,\Delta']\right\rangle_{w,w}&=\int \mathcal D\Delta\,w[\Delta]\mathcal D\Delta'\,w[\Delta']B[\Delta,\Delta']\,.
\end{split}
\label{eq:wAverages}
\end{equation}
The first cumulant is the mean empirical spectral distribution function,
\begin{align}
\kappa_1(x)&=\frac{1}{2}+\frac{c}{2\pi i}\left\langle\Lambda(\Delta'(x),\Delta(x))\right\rangle_{w,w}\nonumber\\
&\quad-\frac{1}{2\pi i}\left\langle\Xi(\Delta(x))\right\rangle_w\,.
\label{eq:kappa1Final}
\end{align}
For the second cumulant, set
\begin{equation}
\begin{split}
\Lambda_{xy}[\Delta,\Delta']&=\Lambda(\Delta'(x),\Delta(x))\\
&\quad\times\Lambda(\Delta'(y),\Delta(y))\,.
\end{split}
\end{equation}
Then the connected two-point cumulant scales as $\kappa_2=O(N^{-1})$ and is
\begin{equation}
\begin{split}
N\kappa_2(x,y)&=\frac{c}{2\pi^2}\left\langle\Lambda_{xy}[\Delta,\Delta']\right\rangle_{w,w}\\
&\quad-\frac{1}{4\pi^2}\left\langle\Xi(\Delta(x))\Xi(\Delta(y))\right\rangle_w\\
&\quad+\frac{1}{4\pi^2}\left\langle\Xi(\Delta(x))\right\rangle_w\left\langle\Xi(\Delta(y))\right\rangle_w \,.
\end{split}
\label{eq:kappa2Final}
\end{equation}
The signs and factors in Eqs.~\eqref{eq:kappa1Final}--\eqref{eq:kappa2Final} follow from the convention in Eq.~\eqref{eq:cgf}; in particular, $\delta^2F/\delta\mu(x)\delta\mu(y)|_{\mu=0}=-N\kappa_2(x,y)$. Higher cumulants can be obtained by taking further functional derivatives, but their explicit expressions are more involved because additional derivatives of the saddle-point distribution $w[\Delta]$ are required; in selected cases, numerical differentiation may provide a simpler alternative~\cite{SM,castillo2018large}.

We now turn to the rate function for Fourier projections of $i_C$. For a fixed interval $[-a,a]$, define
\begin{equation}
g_C=\int dx\,\phi(x)i_C(x)\,,
\label{eq:gCDef}
\end{equation}
where, for the Fourier examples considered below, either $\phi(x)=\cos(n\pi x/a)\chi_{[-a,a]}(x)$ or
$\phi(x)=\sin(n\pi x/a)\chi_{[-a,a]}(x)$, with $\chi_{[-a,a]}$ denoting the characteristic function of $[-a,a]$. These Fourier modes are examples of a broader class of compactly supported test functions $\phi$ to which the same construction applies.

The observable $g_C$ has a direct interpretation as a linear spectral statistic. For compactly supported $\phi$, define
\begin{equation}
\Phi(t)=\int_{-\infty}^{t} ds\,\phi(s)\,.
\label{eq:PhiDefinition}
\end{equation}
For the Fourier examples above, $\Phi(t)=0$ for $t<-a$ and $\Phi(t)=\Phi(a)$ for $t>a$. Using Eq.~\eqref{eq:empiricalDistribution}, one obtains the exact identity
\begin{equation}
g_C=\Phi(a)-\frac{1}{N}\sum_{\alpha=1}^{N}\Phi(\lambda_\alpha(C))\,.
\label{eq:gCLinearStatistic}
\end{equation}
Thus, up to the deterministic shift $\Phi(a)$, $g_C$ is minus a linear statistic of the spectrum. Equivalently, the present formalism gives rate functions for the corresponding class of linear spectral statistics~\cite{diaconis2001linear,guhr1998random,forrester2010loggases}.

The scaled cumulant-generating function of $g_C$ is
\begin{equation}
f(\beta)=-\frac{1}{N}\ln\avC{e^{-N\beta g_C}}\,.
\end{equation}
By Eq.~\eqref{eq:cgf}, it is obtained from the functional $F$ as
\begin{equation}
f(\beta)=F[\beta\phi]\,.
\label{eq:fbetaFbetaPhi}
\end{equation}
With the sign convention of Eq.~\eqref{eq:cgf}, the corresponding candidate rate function $\psi$ is obtained, on a differentiable branch, from
\begin{equation}
u=f'(\beta),\qquad \psi(u)=f(\beta)-\beta u \,.
\label{eq:scalarRateLegendre}
\end{equation}
This is the one-dimensional version of the functional transform in Eq.~\eqref{eq:legendreRate}. Details of the numerical implementation are given in the Supplemental Material~\cite{SM}.

\emph{Comparison with numerical results.---} To validate and illustrate the theory, we compare the replica-symmetric predictions for the mean empirical spectral distribution function $\kappa_1$, the
scaled connected covariance $N\kappa_2$, and a Fourier-projection rate function with exact numerical diagonalization for different values of the mean connectivity $c$.

\begin{figure}
\centering
\includegraphics[scale=0.5]{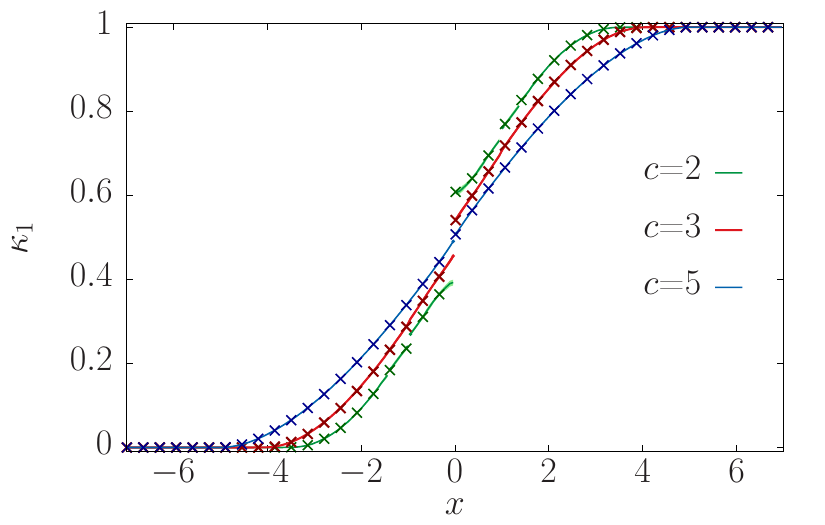}
\caption{Mean empirical spectral distribution function $\kappa_1(x)=\langle i_C(x)\rangle$ for Erd{\H{o}}s--R\'enyi adjacency matrices with mean connectivity $c=2,3,5$. The solid curves are obtained from the population-dynamics solution of Eq.~\eqref{eq:wIf}, while the markers are exact
diagonalization results for $N\times N$ random matrices. The theoretical curves use the interval $[-7,7]$ discretized into $200$ points, regularizer $\eta=10^{-4}$, a population of $10^4$ functions, $10^3$ updates, and $20$ independent population-dynamics runs; the shaded bands show the corresponding run-to-run variability and are visible only for $c=2$. The diagonalization markers are averages over $500$ matrix realizations of size $N=5000$.}
\label{fig:kappa1}
\end{figure}

\begin{figure}
\centering
\includegraphics[scale=0.5]{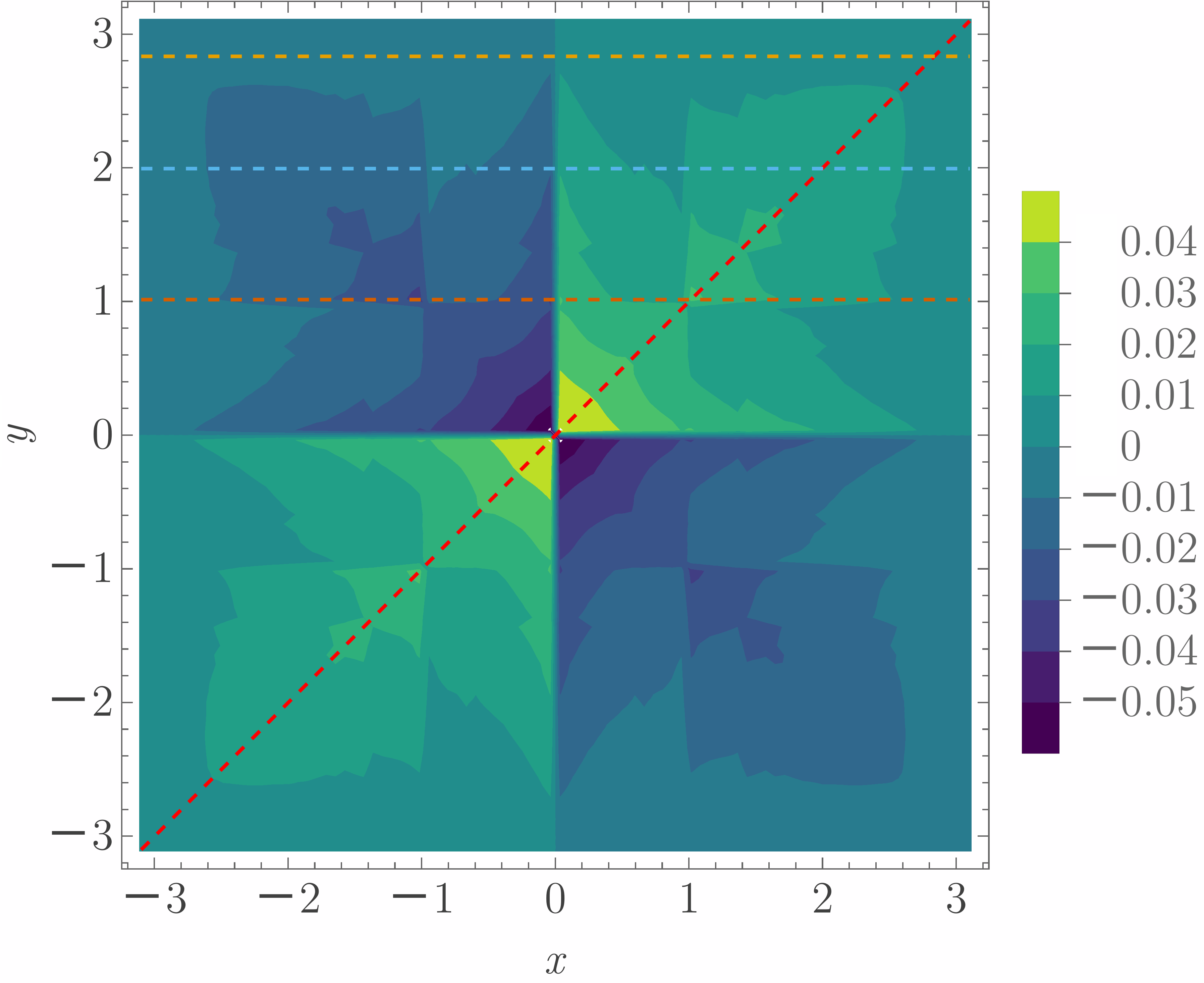}
\begin{minipage}{1\linewidth}
\hspace{-1.3cm}
\includegraphics[scale=0.5]{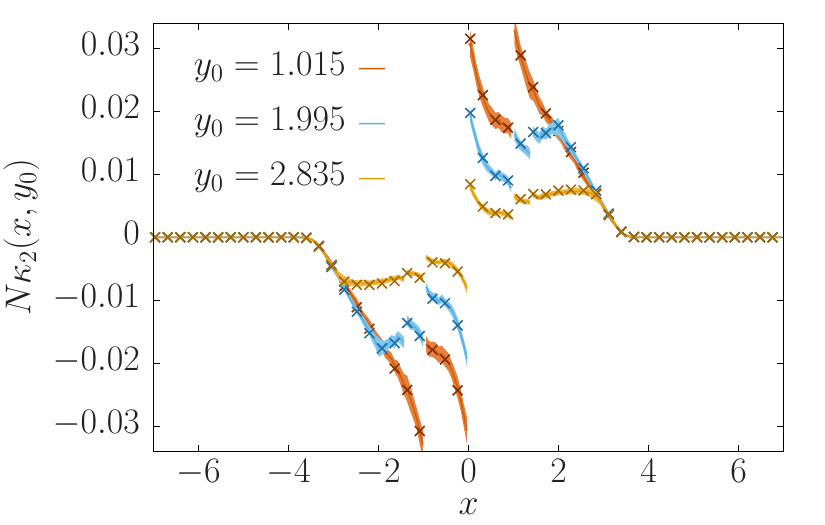}
\end{minipage}
\begin{minipage}{1\linewidth}
\hspace{-1.3cm}
\includegraphics[scale=0.5]{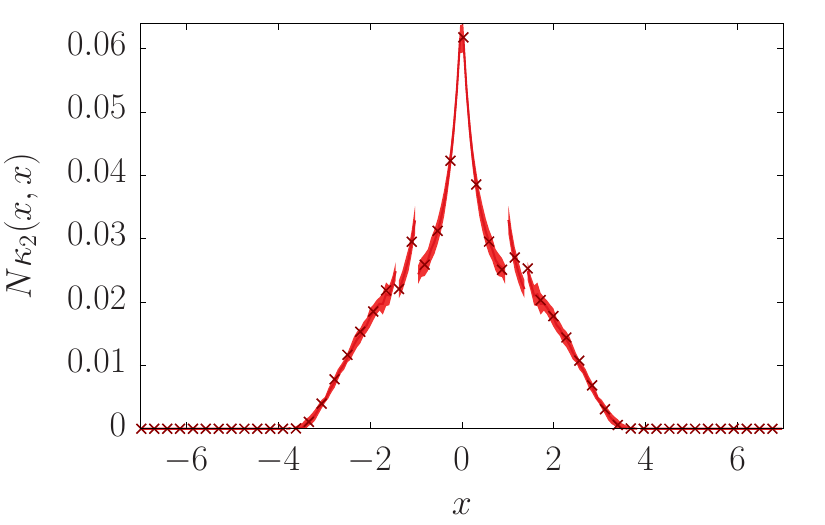}
\end{minipage}
\caption{Scaled connected covariance $N\kappa_2(x,y)$ of the empirical spectral distribution function for Erd{\H{o}}s--R\'enyi adjacency matrices with mean connectivity $c=2$. The top panel shows a density plot of the theoretical population-dynamics result obtained from Eq.~\eqref{eq:wIf}. The middle panel shows cuts at the dashed horizontal lines in the density plot, and the bottom panel shows the diagonal $x=y$. In the cuts, solid curves are the theoretical prediction and markers are exact diagonalization results. The population-dynamics parameters are the interval $[-7,7]$ discretized into $200$ points, regularizer $\eta=10^{-4}$, population size $10^4$, $10^3$ updates, and $20$ independent runs. Shaded bands in the line plots indicate the corresponding run-to-run variability. The diagonalization data are averaged over $500$ matrix realizations of size $N=5000$.}
\label{fig:kappa2}
\end{figure}

Equation~\eqref{eq:wIf} does not generally admit an analytical solution. However, $w$ can be efficiently sampled for arbitrary $\mu$ using a population dynamics algorithm \cite{mezard2001bethe}; explicit details can be found in \cite{kuhn2008spectra, Metz2016, Castillo2018}. The procedure begins by discretizing an interval containing the support of $\mu$, using a resolution sufficient to evaluate the integrals in Eq.~\eqref{eq:wIf} accurately. An
initial population of functions $\{\Delta_\ell\}_{\ell=1}^{\mathcal N}$, defined on this discretization and serving as a preliminary sample of $w$, is iteratively updated according to Eq.~\eqref{eq:wIf} until convergence. The population can then be used to compute averages involving $w$, such as those appearing in Eqs.~\eqref{eq:kappa1Final}--\eqref{eq:kappa2Final}, and to
evaluate $F[\mu]$ for the Fourier-rate calculation below. For Figs.~\ref{fig:kappa1} and \ref{fig:kappa2}, the theoretical calculations use the interval $[-7,7]$ discretized into $200$ points, regularizer $\eta=10^{-4}$, a population of $10^4$ functions, $10^3$ population-dynamics
updates, and $20$ independent population-dynamics runs. The corresponding exact diagonalization data are averaged over $500$ Erd{\H{o}}s--R\'enyi matrices of size $N=5000$.

In Fig.~\ref{fig:kappa1}, we plot $\kappa_1(x)=\langle i_C(x)\rangle$ for several values of $c$, showing good agreement between exact diagonalization and the theoretical prediction across the displayed range of $x$. The observed relation $\kappa_1(-x)=1-\kappa_1(x)$ reflects the spectral symmetry of the locally tree-like cavity problem in the thermodynamic limit \cite{Susca_2021};
it should not be read as an exact finite-$N$ identity for every graph realization. The discontinuity at $x=0$, present for all values of $c$, reflects the delta-peak contribution of zero modes. Additional discontinuities at $c=2$ for $x=\pm1,\pm\sqrt{2}$ originate from small trees with two and three vertices, respectively \cite{Golinelli2003}. As $c$ increases, the weight of these delta contributions decreases, while the support of the continuous spectrum expands.

In Fig.~\ref{fig:kappa2}, we plot the scaled connected covariance $N\kappa_2(x,y)$ for $c=2$, showing several cuts at different values of $y$ as well as the diagonal $x=y$. The discontinuities associated with delta-peak contributions from small trees are more visible than in $\kappa_1$, and they also contribute to the increased run-to-run variability of the theoretical curves. Correlations are generally stronger for values of $y$ near zero and decay gradually along each cut toward the edge of the spectrum. The theoretical curves also reflect the expected thermodynamic symmetry under simultaneous sign inversion, $(x,y)\to(-x,-y)$.

\begin{figure}
\centering
\includegraphics[scale=0.5]{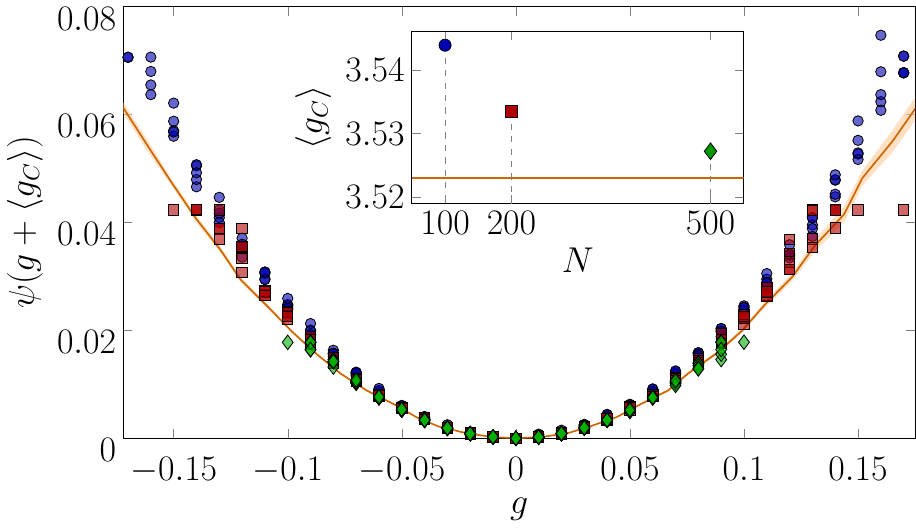}
\caption{Candidate rate function $\psi$ for the Fourier projection $g_C=\int_{-a}^{a}dx\,\sin(\pi x/a)i_C(x)$, with $a=7$, for Erd{\H{o}}s--R\'enyi adjacency matrices with mean connectivity $c=5$. The solid curve is the replica-symmetric prediction. The horizontal coordinate is shifted by the mean, $g=u-\langle g_C\rangle$, so that the minimum is plotted at $g=0$, and the normalization is chosen such that $\psi(\langle g_C\rangle)=0$. The inset shows $\langle g_C\rangle$ as a function of $N$, with the solid horizontal line indicating the theoretical value. Exact diagonalization estimates use $4\times10^4$ samples for each system size and are repeated $5$ times, producing
the scatter points. The theoretical curve was obtained by discretizing the integrals into $80$ subintervals, using a population of $3\times10^4$ functions updated $600$ times; the calculation was repeated $35$ times, and the shaded band shows the resulting run-to-run variability.}
\label{fig:fourierRate}
\end{figure}

In Fig.~\ref{fig:fourierRate}, we plot the candidate rate function of the Fourier projection $g_C=\int_{-a}^{a}dx\,\sin(\pi x/a)\,i_C(x)$ with $a=7$ for $c=5$, along with numerical results for system sizes $N=100$, $200$, and $500$. Smaller systems allow exploration farther from the typical value near the minimum, but finite-size effects limit accuracy in the tails. Larger systems
extend the region of agreement with the theory. Because extreme fluctuations are exponentially rare and challenging to sample directly \cite{Metz2016}, probing the deep tails would require substantially more samples or importance-sampling methods \cite{hartmann2002sampling, hartmann2011sampling}, which we did not employ here. Thus the comparison supports the replica-symmetric prediction in the accessible large-fluctuation regime.

\emph{Concluding Remarks.---} In this work, we developed a replica-based framework for the scaled cumulant-generating functional of the empirical spectral distribution function $i_C$ in diluted Hermitian random matrices. For unweighted adjacency matrices of Erd{\H{o}}s--R\'enyi random graphs at fixed mean degree, the replica-symmetric saddle yields a candidate rate functional, explicit
predictions for the first two cumulants of $i_C$, and rate-function predictions for selected Fourier projections, equivalently selected linear spectral statistics. Exact diagonalization shows good agreement for the mean empirical spectral distribution function and its scaled covariance, and supports the Fourier-rate prediction in the accessible fluctuation regime. Because this comparison relies on direct sampling, it does not probe the very deep tails of the large-deviation regime.

The present formulation has several limitations that also identify natural extensions. The application treated here is the constant-mean-degree, locally tree-like, unweighted Erd{\H{o}}s--R\'enyi ensemble; weighted sparse matrices, heterogeneous degree distributions, and other network ensembles would require the corresponding modifications of the saddle equations. The calculation
assumes a replica-symmetric saddle and a mean-field, locally tree-like structure, and no rigorous large-deviation principle is proved here. Extending the framework to additional spectral observables, weighted or heterogeneous sparse ensembles, geometrically structured or correlated networks, and non-Hermitian matrices are promising directions. The main outcome is a concrete replica-symmetric route to rate-function predictions for functional spectral observables in sparse random matrix ensembles.

\emph{Acknowledgments.---}
This work was supported by the National Natural Science Foundation of China under Grant No. W2511080 and by the Hainan Provincial Natural Science Foundation of China under Grant No. 126MS0008.

\bibliography{references}

\hypersetup{pageanchor=false}
\setcounter{equation}{0}
\setcounter{figure}{0}
\setcounter{table}{0}
\makeatletter
\renewcommand{\theequation}{S\arabic{equation}}
\renewcommand{\thefigure}{S\arabic{figure}}
\renewcommand{\thetable}{S\arabic{table}}
\renewcommand{\thesection}{S\arabic{section}}
\renewcommand{\theHequation}{S\arabic{equation}}
\renewcommand{\theHfigure}{S\arabic{figure}}
\renewcommand{\theHtable}{S\arabic{table}}
\renewcommand{\theHsection}{S\arabic{section}}

\onecolumngrid

\newpage
\makeatletter

\begin{center}
\textbf{\large Supplemental Material for ``\@title''} \\[10pt]
E.{} Guzm\'an-Gonz\'alez$^{1}$ and I.{} P\'erez Castillo$^{2}$
 \\

 \vspace{10pt}

\emph{$^1$
School of Physics and Optoelectronic Engineering, Hainan University,
570228 Haikou, P. R. China
}\\
 \vspace{5pt}

\emph{$^2$
Departamento de F\'isica, Universidad Aut\'onoma Metropolitana-Iztapalapa,
San Rafael Atlixco 186, Ciudad de M\'exico 09340, M\'exico
}\\
\end{center}

\vspace{10pt}

\makeatletter
\setcounter{secnumdepth}{3}
\makeatother

\setcounter{section}{0}
\setcounter{page}{1}
\renewcommand{\thepage}{S\arabic{page}}
\renewcommand{\thesection}{S\arabic{section}}
\makeatletter

\twocolumngrid

In this Supplemental Material we provide:
\begin{enumerate}
\item Details of the procedure used to compute the rate functional of $i_C$ using the replica method.
\item Explicit expressions for the first and second cumulants of $i_C$
 and for its cumulant-generating functional.
\item A stationarity argument showing when the implicit dependence of the functional $w$ on $\mu$ can be omitted in differentiating $F[\mu]$, with applications to $\kappa_1$, $\kappa_2$, and $f'(\beta)$.
\item An explanation of how the rate functional can be used to obtain the candidate rate function of the Fourier coefficients of $i_C$.
\end{enumerate}

\section{Using the replica method to compute the rate functional}
\label{app:asymmetric}
In this section, we present the details to obtain the rate functional using the replica method
in the thermodynamic limit $N \rightarrow \infty$.  We have,
by (\ref{eq:fmulimits}),
\begin{equation}
\begin{split}
F[\mu]&=-\lim_{N \rightarrow \infty}\lim_{\Delta x \rightarrow 0} \lim_{\eta \rightarrow 0^+} \lim_{n_j^{\pm}\rightarrow \pm \Delta x \mu(x_j)/\pi i }\\
& \frac{1}{N}\ln \avC{ \prod_{j=1}^L [\mathcal Z(x_j^\eta)]^{n_j^+}[\mathcal Z^*(x_j^\eta)]^{n_j^-}}\,,
\end{split}
\end{equation}
For the sake of simplicity, in what follows, we stop writing the previous limits. We will also
interchange the order of the limits when necessary, consistency is verified a posteriori
when comparing with numerical diagonalization.

By assuming that $n_j^\pm$ is an integer, we can write the product as a Gaussian integral in a
high dimensional space, and therefore we obtain,
\begin{equation}
\begin{split}
&F[\mu] =\frac{1}{2} \int \mu(x) dx-\frac{1}{N} \ln  \int dy^{n^+ N} dv^{n^- N}\\&
\exp\left(-\frac{i}{2}\sum\limits_{l=1}^L\sum\limits_{n=1}^N\left[\sum\limits_{a=1}^{n_l^+}y^2_{lan} (x^\eta_l )^*-\sum\limits_{b=1}^{n_l^-}v_{lbn}^2x^\eta_l\right]\right)\\&
\avC{\exp\left(\frac{i}{2}\sum\limits_{l=1}^L\sum\limits_{m , n=1}^N \! c_{mn} \left[\sum\limits_{a=1}^{n_l^+}y_{lam} y_{lan}-\sum\limits_{b=1}^{n_l^-}v_{lbm} v_{lbn} \right]
\right)}
\label{eq:FmuInitial}
\end{split}
\end{equation}
where we introduced the replicated variables $y_{lam}, v_{lbm}$, with $l=1, \dots,L$, $a=1, \dots,n_l^+$, $b=1, \dots, n_l^-$, $m=1, \dots,N$, while $dy^{n^+ N}dv^{n^- N}$ indicates their corresponding integration measure.

The following steps depend on the details of the ensemble under consideration. As an illustrative example, we consider the adjacency matrix of Erd{\H{o}}s--R\'enyi random graphs. In this case, $c_{nn}=0$ and $c_{nm}=c_{mn}$ for $m<n$, with these entries independent and distributed as in \eqref{eq:pPoisson}, so that the average in \eqref{eq:FmuInitial} factorizes and can be computed explicitly,
\begin{equation}
\begin{split}
F[\mu] &=\frac{1}{2}\int \mu(x)dx- \frac{1}{N} \ln  \int dy^{n^+ N} dv^{n^- N}\\&
\exp\Bigg\{-\frac{i}{2}\sum\limits_{l=1}^L\sum\limits_{n=1}^N\Bigg[\sum\limits_{a=1}^{n_l^+}y^2_{lan} (x^\eta_l)^*-\sum\limits_{b=1}^{n_l^-}v_{lbn}^2 x^\eta_l \Bigg]\\
&+\frac{c}{N}\sum_{m<n}\!\!\Bigg(e^{i \sum_{l=1}^L\left(\sum_{a=1}^{n_l^+} \!\! y_{lam}  y_{lan}-\sum_{b=1}^{n_l^-} \! \!v_{lbm}  v_{lbn}\right)}-1 \Bigg)\Bigg\}\,,
\end{split}
\end{equation}
where we omitted terms of order $O(N^{-1})$. To decouple the indices $m$ and $n$, we consider a probability density function that plays the role of an \emph{order parameter},
\begin{equation}
\begin{split}
&\tilde P(\bm y,\bm v) = \frac{1}{N} \sum_{j=1}^N \prod_l^L \Bigg[\prod_{a=1}^{n^+_l}\delta (y_{la}-y_{laj})\Bigg]\Bigg[\prod_{b=1}^{n^-_l}\delta(v_{lb}-v_{lbj}) \Bigg]
  \end{split}
\end{equation}
where $\bm y$  denotes a vector with components $y_{la}$, $l=1, \dots,L; a=1, \dots,n^+_l$, while $\bm v$ has components $v_{lb}$, ${l=1, \dots,L; b=1, \dots,n^-_l}$.

By using this expression for $\tilde P$ and omitting terms of order $O(N^{0})$ we can write,
\begin{equation}
\begin{split}
&\frac{c}{N}
\sum_{m<n}\!\!\Bigg(e^{i \sum_{l=1}^L\left(\sum_{a=1}^{n_l^+} \!\! y_{lam}  y_{lan}-\sum_{b=1}^{n_l^-}v_{lbm}  v_{lbn}\right)}-1 \Bigg)=\\
&\frac{cN}{2}\int d \bm y d \bm v  d\bm y'  d\bm v'\tilde P(\bm y,\bm v) \tilde P(\bm y',\bm v')\left(e^{i \left( \bm y \cdot \bm y'-\bm v \cdot \bm v'\right)}-1 \right)\,,
  \end{split}
\end{equation}
where $d \bm y d \bm v$ denotes the volume element in the space of $\bm y$, $\bm v$ and
the dot products are defined in the usual way, $\bm y \cdot \bm y'=\sum_{l=1}^L\sum_{a=1}^{n^+} y_{la} y'_{la}$,  $\bm v \cdot \bm v'=\sum_{l=1}^L\sum_{b=1}^{n^-} v_{lb} v'_{lb}$.

Next, we introduce a functional Dirac delta over the space of the order parameter---the space of all probability distributions $P(\bm y, \bm v)$---which we write explicitly in its Fourier representation, $\delta[P-\tilde P] = \int \mathcal{D} \hat P \, \exp \Big[ i N \int d\bm y  d\bm v \hat P(\bm y, \bm v) \big(P(\bm y, \bm v)-\tilde P(\bm y, \bm v)\big) \Big]$, where $\mathcal{D} \hat P$ denotes a path integral over functions $\hat P(\bm y, \bm v)$. As anticipated, this allows decoupling of the indices $m$ and $n$, giving,
\begin{equation}
\begin{split}
F[\mu]&=\frac{1}{2}\int \mu(x)dx-\frac{1}{N}\ln\int \mathcal D P  \mathcal D\hat P e^{-N\tilde{\mathcal S} (P,\hat P)}\,,
\end{split}
\end{equation}
where $\mathcal D P$ denote a path integral over the space of functions $P(\bm y, \bm v)$, and we defined the action,
\begin{widetext}
\begin{align}
\tilde{\mathcal S} (P,\hat P) &=-\frac{c}{2}\int d \bm y  d\bm y' d \bm v  d\bm v'P(\bm y,\bm v) P(\bm y',\bm v')\left(e^{i \left( \bm y \cdot \bm y'-\bm v \cdot \bm v'\right)}-1 \right)-i  \int d \bm y d \bm v P(\bm y, \bm v) \hat P(\bm y, \bm v)\nonumber\\
&- \ln\left(\int d \bm y  d \bm ve^{-i\hat P(\bm y,\bm v)-\frac{i}{2}\sum_{l=1}^L\big(\sum_{a=1}^{n_l^+}(x^\eta_l)^* y^2_{la}-\sum_{b=1}^{n_l^-}x^\eta_lv^2_{lb}\big )}\right)\,.
\label{eq:actionSIx}
\end{align}
\end{widetext}
The asymptotic behavior of the free energy $F$ for large $N$ can be computed using the saddle-point method by finding the stationary distributions of $\tilde{\mathcal S}$. The corresponding saddle equations are,
\begin{equation*}
\begin{split}
\hat P(\bm y,\bm v)&= ic\int d\bm y'  d\bm v'P(\bm y',\bm v')\left(e^{i \left( \bm y \cdot \bm y'-\bm v \cdot \bm v'\right)}-1\right)\,,\\
P(\bm y,\bm v)&=Ae^{-i  \hat P(\bm y,\bm v)-\frac{i}{2} \sum_{l=1}^L(\sum_{a=1}^{n_l^+}(x^\eta_l)^*y{}^2_{la}-\sum_{b=1}^{n_l^-}x^\eta_l v{}^2_{lb})}\,,
\end{split}
\label{eq:saddleAcc2}
\end{equation*}
where $A$ is a normalization factor that ensures that $P$ is correctly normalized.

\subsection{The replica-symmetric ansatz}
To solve the previous equations, we assume a replica-symmetric ansatz, which is justified by the fact that, on Erd{\H{o}}s--R\'enyi random graphs, replica symmetry is not expected to be broken \cite{mezard1988spin}.

In the replica-symmetric ansatz, we assume the following solution
\begin{equation}
\begin{split}
P(\bm y,\bm v)&= \int d\bm \Delta\, w(\bm \Delta)\prod_{l=1}^L \Bigg\{\Bigg[\prod_{a=1}^{n_l^+}\frac{e^{\frac{i y^2_{la}}{2 \Delta_l}}}{\sqrt{2 \pi i \Delta_l}}\Bigg]\\
&\quad\times\Bigg[ \prod_{b=1}^{n_l^-}\frac{e^{\frac{v^2_{lb}}{2 i \Delta_l^*}}}{\sqrt{-2 \pi i\Delta_l^*}}\Bigg]\Bigg\}\,.
\end{split}
\label{eq:ers1}
\end{equation}
where $\bm \Delta=(\Delta_1, \dots, \Delta_L)$ denotes a complex vector with $L$ entries.

After computing the resulting Gaussian integrals, the first saddle equation reads,
\begin{equation}
\begin{split}
\frac{i\hat P(\bm y,\bm v)}{c}&=1-\int d\bm \Delta w(\bm \Delta)\prod_{l=1}^L \Bigg\{\Bigg[\prod_{a=1}^{n^+_l}e^{\frac{y_{la}^2 \Delta_l}{2i} }\Bigg]\\
&\quad\times\Bigg[ \prod_{b=1}^{n^-_l}e^{\frac{iv_{lb}^2 \Delta_l^*}{2} }\Bigg]\Bigg\}\,.
\end{split}
\end{equation}
Substituting this result in the second saddle equation and performing the remaining Gaussian integrals
yields,
\begin{equation}
\begin{split}
P(\bm y, \bm v)&= A \int d \bm \Delta \sum_{k=0}^\infty p_c(k) \int  [w(\bm \Delta_1)d\bm \Delta_1 \dots w(\bm \Delta_k)d \bm \Delta_k]\\
&\prod_{l=1}^L \delta(\Delta_l-s_{l;\bm \Delta_{1}, \dots ,\bm  \Delta_{k}})(2 \pi i \Delta_l)^{\frac{n_l^+}{2}}(-2 \pi i \Delta_l^*)^{\frac{n_l^-}{2}}\\
&\prod_{l=1}^L\Bigg[\prod_{a=1}^{n_l^+}\frac{e^{\frac{iy^2_{la}}{2 \Delta_l}}}{\sqrt{2 \pi i \Delta_l}} \Bigg]\Bigg[\prod_{b=1}^{n_l^-}\frac{e^{
\frac{v^2_{lb}}{2i\Delta_l^*}}}{\sqrt{-2 \pi i \Delta_l^*}}\Bigg]\,,
  \end{split}
\end{equation}
where we defined $s_{l;\bm \Delta_{1}, \dots ,\bm  \Delta_{k}}  = -(x_l-i\eta+\sum_{r=1}^k \Delta_{rl})^{-1}$, with $\Delta_{rl}$ denoting the $l$-th component of $\bm \Delta_r$. By comparing with (\ref{eq:ers1}) we conclude,
\begin{equation}
\begin{split}
w(\bm \Delta)&=A\sum_{k=0}^\infty p_c(k) \int w(\bm \Delta_1)d\bm \Delta_1 \dots w(\bm \Delta_k)d \bm \Delta_k\\
&\quad\times \prod_{l=1}^L \Bigg[\delta(\Delta_l-s_{l;\bm \Delta_{1}, \dots , \bm \Delta_{k}})\\
&\qquad\qquad\times(2 \pi i \Delta_l)^{\frac{n_l^+}{2}}(-2 \pi i \Delta_l^*)^{\frac{n_l^-}{2}}\Bigg]\,.
  \end{split}
\label{eq:wBefore}
\end{equation}
where, we recall that $A$ is simply a normalization factor.

Taking the continuous limit $\Delta x \to 0$, $\bm \Delta$ becomes a function $\Delta$ defined
over the support of $\mu$, the product of delta functions is replaced by a functional Dirac delta $\delta[\cdot]$, and $w$ becomes a functional of $\Delta$. The previous equation then reads
\begin{equation}
\begin{split}
w[\Delta]&=A\sum_{k=0}^\infty p_c(k) \int [w[ \Delta_1]\mathcal D \Delta_1 \dots w[ \Delta_k] \mathcal D \Delta_k]\\
&\delta[\Delta-s_{\Delta_{1}, \dots , \Delta_{k}}]e^{\frac{1}{2 \pi i}\int dx  \mu(x)\Xi(\Delta(x) )}
\end{split}
\label{eq:wASaddle}
\end{equation}
where we used the fact that, in this limit, $s_{l;\bm \Delta_{1}, \dots ,\bm  \Delta_{k}}$ goes to the function defined in \eqref{eq:sDeltaDef}, and,
\begin{equation}
\begin{split}
\lim_{\Delta x \rightarrow 0}\prod_{l=1}^L(2 \pi i \Delta_l)^{\frac{n_l^+}{2}}(-2 \pi i \Delta_l^*)^{\frac{n_l^-}{2}}=e^{\frac{1}{2 \pi i}\int dx  \mu(x) \Xi(\Delta(x) )}\,,
\end{split}
\label{eq:for3209}
\end{equation}
where $\Xi$ is defined in equation (\ref{eq:functionDefinitions}).

Finally, we rewrite the action $\tilde{\mathcal S}$ (\ref{eq:actionSIx}). After applying the replica-symmetric ansatz, computing the resulting Gaussian integrals, and taking the continuous limit, we obtain
\begin{equation}
\begin{split}
\tilde{\mathcal S} &=- \frac{c}{2}+  \frac{c}{2}\int \mathcal D \Delta\mathcal D \Delta'  w[\Delta] w[\Delta']e^{\frac{1}{\pi i}\int dx \mu(x) \Lambda(\Delta'(x),\Delta(x) )}\\
&- \ln \Bigg\{\sum_{k=0}^\infty p_c(k)\int \mathcal D \Delta  \mathcal D\Delta_1 w[\Delta_1]\dots\mathcal D \Delta_kw[\Delta_k]\\
&\qquad \qquad \times \delta[\Delta-s_{\Delta_{1}, \dots , \Delta_{k}}]e^{\frac{1}{2 \pi i}\int dx  \mu(x)\Xi(\Delta(x) )}\Bigg\}\,,
\end{split}
\label{eq:SOmegaFinal}
\end{equation}
where $\Lambda$ and $\Xi$ are defined in \eqref{eq:functionDefinitions}. Equations \eqref{eq:FmuC} and \eqref{eq:SmuC} then follow immediately. We emphasize that the previous equation holds without assuming (\ref{eq:wASaddle}); it relies only on the replica-symmetric ansatz.

\section{Computing the first two cumulants and the rate functional of $i_C$}
\label{app:asymmetricCumulants}
As shown in \eqref{eq:FmuC}, the cumulants of the function $i_C$ can be obtained by computing functional derivatives of $F[\mu]$ and then setting $\mu \equiv 0$. This computation must take into account that $w$ depends on $\mu$ via equation \eqref{eq:wIf}. As shown in Sec.~\ref{sec:implicitDependence}, stationarity of the saddle-point functional implies that this implicit dependence does not contribute to the first functional derivative for arbitrary $\mu$. For the second derivative used to compute $\kappa_2$, the cancellation is established at $\mu\equiv0$, which is the only case needed for the connected covariance. We proceed to derive these expressions.

If we compute the functional derivatives of $F[\mu]$ and then set $\mu \equiv 0$, we obtain
\begin{equation}
\begin{split}
\exValue{i_C(x)}&=\frac{1}{2} +\frac{c}{2 \pi i} \int \mathcal D  \Delta w[\Delta] \mathcal D  \Delta' w[\Delta']\Lambda(\Delta'(x), \Delta(x) )\\
&-\frac{1}{2 \pi i}\int \mathcal D  \Delta w[\Delta]  \Xi(\Delta(x) )
\end{split}
\end{equation}
where we used that, for a general functional $G$ we can write,
\begin{equation}
\begin{split}
&\sum_{k=0}^\infty p_c(k)\int  w[\Delta_1]\mathcal D\Delta_1 \dots w[\Delta_k] \mathcal D \Delta_kG[s_{\Delta_1, \dots, \Delta_k}]\\
&=\frac{\int \mathcal D  \Delta w[\Delta] G[\Delta]e^{-\frac{1}{2 \pi i}\int dx  \mu(x)\Xi( \Delta(x) )}}{\int \mathcal D  \Delta w[\Delta]e^{-\frac{1}{2 \pi i}\int dx  \mu(x) \Xi (\Delta(x) )}}\,,
\end{split}
\label{eq:usefulFormula}
\end{equation}
as can be verified by considering (\ref{eq:wASaddle}).

Similarly, computing the second functional derivatives of $F[\mu]$ yields the connected correlation function $\kappa_2(x,y) = \exValue{i_C(x) i_C(y)}-\exValue{i_C(x)}\exValue{i_C(y)}$,
\begin{equation}
\begin{split}
N\kappa_2(x,y)&=\frac{c}{2 \pi^2 } \int \mathcal D \Delta w[\Delta] \mathcal D \Delta' w[\Delta']\Lambda_{xy}[\Delta, \Delta']\\
&-\frac{1}{4 \pi^2}\int \mathcal D \Delta w[\Delta]\Xi(\Delta(x) )\Xi(\Delta(y) )\\
&+\frac{1}{4 \pi^2}\left(\int \mathcal D \Delta w[\Delta]\Xi(\Delta(x) )\right)\\
&\quad\times\left(\int \mathcal D \Delta w[\Delta]\Xi(\Delta(y) )\right)\,,
\end{split}
\end{equation}
where $\Lambda_{xy}[\Delta, \Delta']= \Lambda(\Delta'(x), \Delta(x) )\Lambda(\Delta'(y), \Delta(y) )$.

Finally, using (\ref{eq:usefulFormula}), we obtain the following simplified expression for the extremized functional,
\begin{equation}
\begin{split}
F[\mu] &= \frac{1}{2} \int \mu(x) dx-\frac{c}{2}\\
&+\ln \int \mathcal D  \Delta w[\Delta]e^{-\frac{1}{2 \pi i}\int dx  \mu(x) \Xi (\Delta(x) )}\\
&+\frac{c}{2} \int \mathcal D  \Delta w[\Delta] \mathcal D  \Delta' w[\Delta']e^{\frac{1}{\pi i}\int dx \mu(x) \Lambda(\Delta'(x),\Delta(x) )}\,.
\end{split}
\label{eq:FmuFinal}
\end{equation}
Since this expression is valid for any $\mu$, it is useful when computing the candidate rate function using (\ref{eq:legendreRate}).

\section{On the Implicit Dependence of $w$ for the First Two Cumulants}
\label{sec:implicitDependence}
We clarify why the implicit dependence of the saddle-point functional $w$ on $\mu$ can be omitted in the derivatives used above. For the first derivative, the cancellation follows from stationarity and holds for arbitrary $\mu$. For the second derivative, the argument below shows the required cancellation at $\mu(x)\equiv0$, which is sufficient for the computation of $\kappa_2$.

The starting point is equation (\ref{eq:actionSIx}) for the action $\tilde{\mathcal S} (P, \hat P)$.
In order to evaluate $F[\mu]$, $\tilde{\mathcal S} (P, \hat P)$ needs to be evaluated at the $P$ and $\hat P$ that extremize it after fixing $\mu$. Note that if we write
\begin{widetext}
\begin{equation}
\begin{split}
P(\bm y,\bm v)&= \int d\bm \Delta\, w(\bm \Delta)\prod_{l=1}^L \Bigg[\prod_{a=1}^{n_l^+}\frac{e^{\frac{i y^2_{la}}{2 \Delta_l}}}{\sqrt{2 \pi i \Delta_l}}\Bigg]\Bigg[ \prod_{b=1}^{n_l^-}\frac{e^{\frac{v^2_{lb}}{2 i \Delta_l^*}}}{\sqrt{-2 \pi i\Delta_l^*}}\Bigg]\,,\\
\frac{i\hat P(\bm y,\bm v)}{c}&=1-\int d\bm \Delta w(\bm \Delta)\prod_{l=1}^L\Bigg[\prod_{a=1}^{n^+_l}e^{\frac{y_{la}^2 \Delta_l}{2i} }\Bigg]\Bigg[ \prod_{b=1}^{n^-_l}e^{\frac{iv_{lb}^2 \Delta_l^*}{2} }\Bigg]\,.
  \end{split}
  \label{eq:omegaEmbedding}
\end{equation}
\end{widetext}
with $w$ an arbitrary distribution over the space of vectors $\bm \Delta$,  then, after taking the
continuous limit, the action can be written as in equation \eqref{eq:SOmegaFinal}, without the need to assume \eqref{eq:wASaddle}. We can then define an effective action, $S_{\text{eff}}[\mu, w]$, acting on the space of functionals $w$. Note that this space can be embedded  in the larger space of functions $P$ and $\hat P$ via \eqref{eq:omegaEmbedding}. This means that any extremum of $\tilde{\mathcal S} (P, \hat P)$ that lies in the smaller space  is an extremum $S_{\text{eff}}[\mu, w]$, but not the other way around. An extremum of $S_{\text{eff}}[\mu, w]$ is not necessarily an extremum of $\tilde{\mathcal S} (P, \hat P)$, there might be directions outside the embedded space where the action is not stationary.

Given a functional $w^*$ that satisfies \eqref{eq:wASaddle}, by construction, it extremizes $\tilde{\mathcal S} (P, \hat P)$, with $P$ and $\hat P$ given by \eqref{eq:omegaEmbedding}. Therefore, for all functions $\Delta$ and $\mu$ the following result holds
\begin{equation}
\begin{split}
0=\frac{ \delta S_{\text{eff}}}{ \delta w[ \Delta]} [\mu,w^*]\,.
\end{split}
\label{eq:SeffOmega}
\end{equation}
Alternatively, it is a simple algebraic exercise to verify that if $w^*$ satisfies \eqref{eq:wASaddle}, then  \eqref{eq:SeffOmega} holds.

In terms of this effective action, the functional $F$ is
\begin{equation}
\begin{split}
F[\mu]= S_{\text{eff}}[\mu, w^*]\,,
\end{split}
\end{equation}
where $w^*$ depends on $\mu$ via \eqref{eq:wASaddle}. The functional derivative is therefore,
\begin{equation}
\begin{split}
\frac{\delta F}{\delta \mu(x)}[\mu ] &=\frac{\delta S _{\text{eff}}}{\delta \mu(x)}[\mu, w^*]+ \int d \Delta \frac{\delta  S_{\text{eff}}}{\delta w[\Delta]}[\mu, w^*]\frac{\delta w[\Delta]}{\delta \mu(x)}[w^*]\\
&=\frac{\delta S _{\text{eff}}}{\delta \mu(x)}[\mu, w^*]\,,  
\end{split}
\label{eq:Fmu1}
\end{equation}
where we used \eqref{eq:SeffOmega}. This implies that the dependence of $w^*$ on $\mu$ can be omitted
when taking the first functional derivative of the extremized action; this is the envelope-theorem step used for $\kappa_1$. We now show the corresponding statement needed for $\kappa_2$ at $\mu\equiv0$.

The key observation to continue with the computation is that \eqref{eq:Fmu1}  holds for arbitrary $\mu$. Therefore, by taking another functional derivative we conclude,
\begin{equation}
\begin{split}
\frac{\delta^2 F}{\delta \mu(y)\delta \mu(x)}[\mu] &=\frac{\delta^2 S _{\text{eff}}}{\delta \mu(y)\delta \mu(x)}[\mu, w^*]\\
&+\int d \Delta \frac{\delta^2  S_{\text{eff}}}{\delta w[\Delta]\delta \mu(x)}[\mu, w^*] \frac{\delta w[\Delta]}{\delta \mu(y)}[w^*]\\
&=\frac{\delta^2 S _{\text{eff}}}{\delta \mu(y)\delta \mu(x)}[\mu, w^*]\\
&+\int d \Delta \frac{\delta^2  S_{\text{eff}}}{\delta \mu(x)\delta w[\Delta]}[\mu, w^*] \frac{\delta w[\Delta]}{\delta \mu(y)}[w^*]\,,
\end{split}
\label{eq:dFdmudmu}
\end{equation}
where we only changed the order of the derivatives to obtain the second equality. Since \eqref{eq:SeffOmega} also holds for arbitrary $\mu$,
\begin{equation}
\begin{split}
0&=\frac{ \delta^2 S_{\text{eff}}}{ \delta \mu(x) \delta w[ \Delta]} [\mu,w^*]\\&
+\int d\Delta' \frac{ \delta^2 S_{\text{eff}}}{\delta w[ \Delta']  \delta w[ \Delta]}[\mu, w^*]\frac{\delta w[ \Delta']}{\delta \mu(x)}[w^*]\,.
\end{split}
\label{eq:SeffOmega1}
\end{equation}
The previous equality holds in particular for $\mu\equiv 0$. Note that, in such case,
\begin{equation}
\begin{split}
&S_{\text{eff}}[\mu\equiv0,w]=- \frac{c}{2}+  \frac{c}{2}\int \mathcal D \Delta\mathcal D \Delta' w[\Delta] w[\Delta']\\
&-\ln \Bigg\{\sum_{k=0}^\infty p_c(k)\int \mathcal D \Delta  \mathcal D\Delta_1 w[\Delta_1]\dots \mathcal D \Delta_kw[\Delta_k]\\
&\qquad \qquad \times \delta[\Delta-s_{\Delta_{1}, \dots , \Delta_{k}}]\Bigg\}\\
&=-\frac{c}{2}+\frac{c}{2}+\ln 1=0\,,
\end{split}
\end{equation}
where we used $\int \mathcal D\Delta\, w[\Delta]=1$. This implies that $S_{\text{eff}}[\mu\equiv0,w]$ is independent of $w$ (note however, this does not imply that, when $\mu \equiv 0$, the complete action $\tilde{\mathcal S} (P, \hat P)$ is extremized by any $w$,  $P$ and $\hat P$ given by (\ref{eq:omegaEmbedding})), so $\delta^2 S_{\text{eff}}/(\delta w [\Delta] \delta w [\Delta'] )\,[\mu\equiv0,w^*]=0$. By considering this result in (\ref{eq:SeffOmega1}), $0= \delta^2 S_{\text{eff}}/ (\delta \mu(x) \delta w[ \Delta] )\, [\mu\equiv 0,w^*]$. Finally, this result together with (\ref{eq:dFdmudmu}) implies,
\begin{equation}
\begin{split}
\frac{\delta^2 F}{\delta \mu(y)\delta \mu(x)}[\mu \equiv 0] =\frac{\delta^2 S _{\text{eff}}}{\delta \mu(y)\delta \mu(x)}[\mu \equiv 0, w^*]\,.
\end{split}
\end{equation}
Thus, at $\mu\equiv0$, $\kappa_2$ can be computed by considering only the explicit functional derivatives with respect to $\mu$ and discarding the implicit dependence of $w^*$ on $\mu$. Since the previous equality was obtained only at $\mu\equiv0$, it should not be used to compute higher functional derivatives without additional terms.

\section{Computing the candidate rate function for the Fourier coefficients of $i_C$}
\label{sec:FourierCoefficients}
We now describe how to compute the candidate rate function of the random variable $g_C$ defined in Eq.~\eqref{eq:gCDef}.

We begin by computing its cumulant-generating function,
\begin{equation}
\begin{split}
f(\beta)= -\frac{1}{N} \ln \exValue{e^{-N \beta \int dx \phi(x) i_C(x) }}=F[\beta \phi]\,,
\end{split}
\label{eq:fvsF}
\end{equation}
where $F$ denotes the cumulant-generating functional (\ref{eq:cgf}). Assuming the relevant convexity and differentiability conditions hold, by a Legendre transform, we can compute the candidate rate function $\psi$ for $g_C$.

By definition, we know that
\begin{equation}
\begin{split}
\psi(u)= f(\beta)- \beta u
\end{split}
\end{equation}
where $\beta$ is given by,
\begin{equation}
\begin{split}
f'(\beta) = u\,.
\end{split}
\end{equation}
Since the previous equation is not simple to solve in general, instead we take $\beta \in \mathbb{R}$ and compute $u=f'(\beta)$ to conclude,
\begin{equation}
\begin{split}
\psi(f'(\beta) ) = f(\beta)- \beta f'(\beta)\,.
\end{split}
\end{equation}
By taking different values of $\beta$ and considering points of the form $(f'(\beta),\psi(f'(\beta) ) )$ we can plot $\psi$ parametrically.

To connect with the rate functional, we use (\ref{eq:fvsF}) to conclude,
\begin{equation}
\begin{split}
f'(\beta) &= \int dx \frac{\delta F}{\delta \mu(x)}[\beta \phi]\, \frac{d \mu(x)}{d\beta} \Bigl|_{\mu(x)=\beta \phi(x)}\\
&=\int dx \frac{\delta F}{\delta \mu(x)}[\beta \phi]\,  \phi(x)\,.
\end{split}
\end{equation}
Computing the functional derivative of \eqref{eq:FmuC}, followed by \eqref{eq:usefulFormula}, gives the following expression. The implicit dependence of $w$ on $\mu$ can be omitted here because $f'(\beta)$ is a first derivative of the extremized functional evaluated at the source $\mu(x)=\beta\phi(x)$; this is precisely the stationarity argument in Eq.~\eqref{eq:Fmu1}.
\begin{equation*}
\begin{split}
\frac{\delta F}{\delta \mu(x)}[\mu]&=\frac{c}{2\pi i}\int \mathcal D \Delta  w[\Delta]\mathcal D\Delta' w[\Delta']\Lambda(\Delta'(x),\Delta(x) )\\
&\times e^{\frac{1}{\pi i}\int dx' \mu(x') \Lambda(\Delta'(x'),\Delta(x') )}\\
&-\frac{1}{2\pi i}\int \mathcal D \Delta  w[\Delta] \Xi(\Delta(x) )+\frac{1}{2}\,.
\end{split}
\end{equation*}
Finally, $F[\beta \phi]$ can be computed using \eqref{eq:FmuFinal}.

\end{document}